\documentclass[a4paper]{jpconf1}
\usepackage{graphicx}
\usepackage{epsfig, amssymb, amsmath, multicol}
\def\aj{AJ}
\def\araa{ARA\&A}%
\def\apj{ApJ}%
\def\apjl{ApJ}%
\def\ao{Appl.~Opt.}%
\def\aap{A\&A}%
\def\aaps{A\&AS}%
\def\mnras{MNRAS}%
\def\nat{Nature}%

\def\mathstacksym#1#2#3#4#5{\def#1{\mathrel{\hbox to 0pt{\lower#5\hbox{#3}\hss} \raise #4\hbox{#2}}}}

\mathstacksym\gta{$>$}{$\sim$}{1.5pt}{3.5pt} 
\mathstacksym\lta{$<$}{$\sim$}{1.5pt}{3.5pt} 

\begin{document}
\title{New Astrophysical Opportunities Exploiting Spatio-Temporal Optical Correlations}

\author{C. Barbieri$^1$, M.~K.~Daniel$^2$, W.~J.~de Wit$^3$, D.~Dravins$^4$, 
H. Jensen$^4$, P. Kervella$^{5}$, S.~Le Bohec$^6$,  F.~Malbet$^{7,8}$, P.~Nunez$^6$,  
J.~P.~Ralston$^9$, E.~N.~Ribak$^{10}$ {\it for the working group on stellar intensity
  interferometry (IAU commission 54)}}

\address{$^1$ University of Padova, Italy; $^2$ Durham University, UK; $^3$ University of Leeds, UK;\newline
$^4$ Lund Observatory, Sweden; $^5$ LESIA, France; $^6$ University of Utah, USA; $^7$ LAOG, France;\newline
$^8$ Caltech, USA; $^9$ University of Kansas, USA; $^{10}$ Technion, Israel}


\begin{abstract}
The space-time correlations of streams of photons can provide fundamentally
new channels of information about the Universe. Today's astronomical
observations essentially measure certain amplitude coherence functions
produced by a source. The spatial correlations of wave fields has
traditionally been exploited in Michelson-style amplitude interferometry.
However the technology of the past was largely incapable of fine timing
resolution and recording multiple beams.  When time and space correlations
are combined it is possible to achieve spectacular measurements that are
impossible by any other means.  Stellar intensity interferometry $(SII)$ is
ripe for development and is one of the few unexploited mechanisms to obtain
potentially revolutionary new information in astronomy. As we discuss below,
the modern use of $SII$ can yield unprecedented measures of stellar diameters,
binary stars, distance measures including Cepheids, rapidly rotating stars,
pulsating stars, and short-time scale fluctuations that have never been
measured before.
\end{abstract}

\section{Introduction}
More than 40 years ago the basic principles of {\it intensity interferometry}
were developed and proven through the measurement of stellar diameters with
sub-milli-arcsecond resolution \cite{1956Natur.178.1046H,1958RSPSA.248..199B}.  Yet this breakthrough ran its course due to
technological limitations. The old technology was limited to two telescope
beams, and integration of correlations was done by the elegantly primitive
device of counting the turns of an electric motor. In the interim the
technology of photon detection has exploded, and the ability to handle and
correlate multiple beams with exquisite time resolution has been perfected.
It is hard to overstate the advantages now possible with current and future
detection technology, both in real-time and off-line, both with simple pairs
and with multiple detector arrays.

$SII$ works by comparing the random intensity fluctuations of light waves
entering separated detectors.  Nothing special is needed from the source, and
black-body, thermal correlations are ideal.  While quantum properties
derived from the statistics of photon arrival times (e.g. photon bunching
behavior) add extra information, the effect is robust, and works whether or
not quanta are well-resolved. Statistical correlations of beams of a given
wavelength $\lambda$ separated by a given distance $D$ allows resolution of
structure at the Rayleigh criterion $\Delta \theta \sim \lambda/D$. This is how
measurements equivalent to 100 meter telescopes were made with $SII$ 40 years 
ago \cite{1974MNRAS.167..121H}.  The
exacting match of optical path lengths needed in amplitude (Michelson)
interferometry is completely unnecessary, because different phase properties
dominate $SII$. Troublesome sensitivity to path length differences caused by
atmospheric turbulence is completely eliminated for the same reason. 
Vast quantities of photons are however required and the visual 
wavelength region is most appropriate. Several algorithms are available for phase
recovery in a multi-beam set-up, permitting model independent imaging \cite{1978ApOpt..17.2047S,
1998TJPh...22..949V}. Current detection technology allow therefore a revolution in
$SII$: multiple telescope, long baseline optical {\it intensity interferometry}. A
system can be devised \cite{2006MNRAS.368.1652O,2006MNRAS.368.1646O} that
would consist of hundreds of large flux gathering surfaces spread out over kilometers 
of baseline that would enable to make the next big step in astronomy: optical imaging
at $\mu$-arcsecond resolution.



\section{Astrophysical applications for $\mu$-arcsecond imaging $SII$}
\label{SciencePotential}
Properly imaging stellar surfaces constitutes a major break-through in
stellar astrophysics. Stars have angular sizes of tens of
milli-arcseconds or less, and until now, apart from the sun, only
Betelgeuse and Altair have been imaged albeit with a modest number of
resolution elements (at maximum ten). Technical requirements for $SII$
are such that one can easily be looking at hundred resolution elements
with an order of magnitude increase in angular resolution.  Stellar
physics has been waiting for a long time to make this leap forward,
and it is similar to the impact realized by Hanbury Brown and
Twiss when they measured 32 stellar diameters with the first and only
application of $SII$ in astronomy \cite{1974MNRAS.167..121H}. Modern $SII$ would undoubtedly extend
solar physics to the realm of the stars. Moreover, feasibility studies
indicate that $SII$ may not be limited to resolving Galactic sources
only. We present therefore an overview of core galactic and
extra-galactic science drivers:
\begin{itemize}
\item Stellar surface phenomena and dynamo action
\item Conditions for planet formation around young stars
\item Cepheid properties and the distance scale
\item Mass loss and fast rotation of massive hot stars and supernova
progenitors
\item Nuclear optical emission from AGN
\item Structure of gamma-ray bursts
\end{itemize}
Of course, feasibility of these science topics depends on the design of
the intensity interferometer. For practical purposes we adopt
a conservative limiting visual magnitude $\rm m_{v}<8^{\rm m}$ and a
resolution of a 0.1 milli-arcsecond, and we detail the
science that would be opened up within these limits.

\begin{figure*}
\centerline{\includegraphics[height=10.2cm,width=11.5cm]{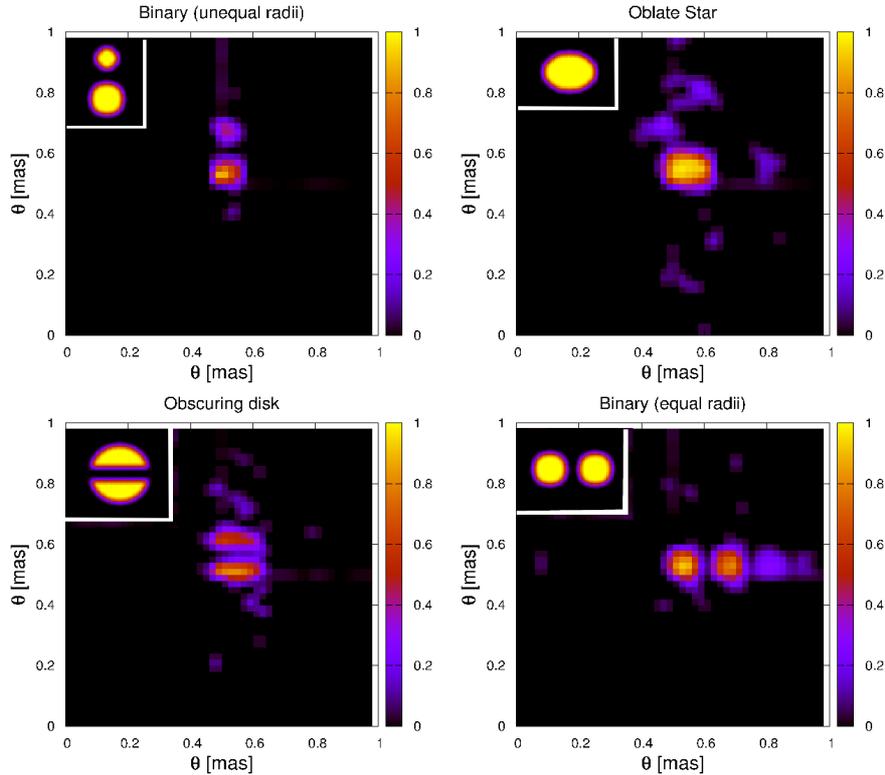}}
  \caption{Four examples of images reconstructed from $SII$ data as would be
recorded with an array that operates at a wavelength of $\sim 400$\,nm with
one hundred telescopes and a telescope separation of $\sim 100$\,m. The pristine image is shown
at the top left in each example. The images  were produced \cite{nunez} using
an algorithm based on the Cauchy-Riemann equations
\cite{2004JOSAA..21..697H}.  The analysis does not yet include a realistic noise
component, which is still actively being investigated.}
  \label{siiimg}
\end{figure*}

\paragraph{\bf{Stellar surface phenomena, star formation and binaries}}
Pre-main sequence (PMS) objects are young stars that are contracting
towards the main-sequence, still lacking internal hydrogen
combustion. Key questions relating to the physics of mass accretion
and PMS evolution can be addressed by means of very high resolution
imaging.  Spatially resolved studies would involve the absolute
calibration of PMS tracks, the mass accretion process, continuum
emission variability, and stellar magnetic activity. The technique may
allow spot features on the stellar surface to be resolved.  Hot spots
deliver direct information regarding the accretion of material onto
the stellar surface. Cool spots, on the other hand, may cover 50\% of
the stellar surface, and they are the product of the slowly decaying
rapid rotation of young stars. Imaging them will constrain 
the interplay of rotation, convection, and chromospheric
activity as traced by cool spots and need not be limited to PMS stars
only. It may also provide direct practical application as the explanation 
for the anomalous photometry observed in young stars \cite{2003AJ....126..833S}.

In the last decade several young coeval stellar groups have been
discovered in close proximity ($\sim$50\,pc) to the sun. Famous
examples are the TW Hydra and $\beta$ Pic comoving groups. The majority of the spectral types
within reach range between A and G-type, and about 50 young stars have $\rm
m_{v}<8^{\rm m}$.  Their ages lie within the range 8
to 50 Myr (see \cite{2004ARA&A..42..685Z} for an overview).  The age
intervals ensures that a substantial fraction of the low-mass members are still
in their PMS contraction phase. Key targets for calibration of evolutionary PMS 
tracks in the Hertzsprung-Russell diagram are spectroscopic binaries. Resolving
binaries delivers the inclination of the system and hence access to the 
mass of the components. This is a fundamental exercise not only for
PMS stars but for all spectroscopic binaries in any evolutionary stage. 
Measurement of angular sizes of individual PMS stars in combination with
a distance estimate (e.g. GAIA) allows a direct comparison between predicted and
observed sizes of these gravitationally contracting stars. The proximity
of the comoving groups ensures that their members are bright. Their proximity renders the
comoving group also relatively sparse making them very suitable, unconfused targets.
The sparseness is also the reason for incomplete group memberships, making it likely that the number of
young stars close to the sun will increase with the years to come.



\paragraph{\bf{Distance scale and pulsating stars}}
Measuring diameters of Cepheids is a basic method with far reaching
implications for the calibration of the distance scale. A radius
estimate of a Cepheid can be obtained using the Baade-Wesselink
method. The Baade-Wesselink method relies on the measurement of the
ratio of the star size at times $t_1$ and $t_2$, based on the
luminosity and color. Combined with a simultaneous measurement of the
radial velocity, this method delivers the difference in the radius
between $t_1$ and $t_2$. With the known difference and ratio of the
radius at two times, one can derive the radius of the
Cepheid. Combining $SII$ angular size measurement with the radius
estimate one obtains the distance (see
\cite{1994ApJ...432..367S}). This makes possible the calibration of
the all important Cepheid period-luminosity relation using local
Cepheids. Nearly all of the {\it Hipparcos} distances for Cepheids
have rather large trigonometric errors (see
\cite{1997MNRAS.286L...1F}) giving rise to ambiguous results. A count
of Cepheids observed with {\it Hipparcos} \cite{1999A&AS..139..245G}
shows that at least 60 Cepheids are visually brighter than $8^{m}$.

\paragraph{\bf {Rapidly rotating stars and hot stars}}
As a group, classical Be stars are particularly well-known for their
close to break-up rotational velocities as deduced from photospheric
absorption lines.  In addition the stars show Balmer line emission
firmly established as due to gaseous circumstellar disks, that appear
and disappear on timescales of months to years. Photometric
observations of Be star disks seem to indicate that they may actually
evolve into ring structures before disappearing into the interstellar
medium (e.g. \cite{2006A&A...456.1027D}). These two phenomena (rapid
rotation and circumstellar disks) are somehow related, but many open
questions exist regarding the detailed physical processes at play.
The Be-phenomenon is important
given the large number of stars and fundamental stellar physics
involved (fraction of Be stars to normal B-type peaks at nearly 50\%
for B0 stars, \cite{1997A&A...318..443Z}). Absorption line studies cannot
provide the final answer to their actual rotational velocity due to
strong gravity darkening at the equator and brightening at the pole
areas. Direct measurement of the shape of the rotating star is not
hampered by gravity darkening, and can provide a direct indication of
the rotational speed (see $\alpha$\,Eri with the VLTI,
\cite{2003A&A...407L..47D}).  The disk's Bremsstrahlung can
constitute $\sim 30\%$ of the total light in $V$-band
\cite{1997A&A...318..443Z}.   There are about 300 Be
stars\footnote{\tt{http://www.astrosurf.com/buil/us/becat.htm}}
brighter than $\rm m_{v}=8^m$, roughly corresponding to a distance
limit of 700\,pc.

The signal-to-noise ratio of an optical intensity interferometer improves with
the temperature of the target. Since $SII$ is insensitive to atmospheric
turbulence, observations in the violet or blue pose no problem. Eliminating
the ``Achilles' Heel'' of standard phase interferometry will enable high resolution studies of hot
stars over very long geometric baselines. A rough estimate of the effective temperatures of the objects in the
Bright Star Catalog reveals that approximately 2600 stars brighter
than visual magnitude 7 and hotter than 9000\,K exist in the sky, all
of which should be realistic targets even for a fairly modest
intensity interferometer. Typical angular sizes for the stellar surface disk
range between 0.5 and 5 mas. Of these targets, a handful of especially
interesting ones are presented below, along with some brief notes on
their possible science potential.

\begin{itemize}

\item {\bf Rigel ($\beta$ Orionis)} Rigel is
the nearest blue supergiant (240\,pc).  It is a very dynamic object with variable
absorption/emission lines and oscillations on many different
timescales - from minutes to weeks \cite{2009AAS...21340813S}. The physical
properties of Rigel have recently been shown to be very similar to 
the supernova progenitor SN1987A. This realization makes understanding
the dynamical nature of Rigel highly relevant.

\item {\bf $\beta$ Centauri} $\beta$
Centauri is a binary system consisting of two very hot, very
massive variable stars. Both components are variable both spatially
and in their line profiles. The intricate nature of the system has led to it being called "a
challenge for current evolution scenarios in close binaries" \cite{2002A&A...384..209A}.

\item {\bf Vega ($\alpha$ Lyrae)} While one of the most fundamental stars in the
sky for calibration purposes, the nature of Vega has proven to be more complicated than
previously thought. Recent phase interferometry studies suggest an
18-fold drop in intensity at 500 nm from center to limb, consistent
with a rapidly rotating pole-on model, as contrasted to the 5-fold
drop predicted by non-rotating models \cite{2006Natur.440..896P}.
Intensity interferometry will direct studies of the intensity distribution at shorter wavelengths
and in different passbands. Such observations will lead to a better
understanding between the link between rotation and limb darkening in
different wavelength regions.

\item {\bf $\eta$ Carinae} $\eta$ Carinae is the most luminous star known in
the Galaxy, and an extremely unstable and complex object. Recent VLTI studies
have revealed asymmetries in the stellar winds due to the rapid rotation of
the star \cite{2007A&A...464...87W}. Like Rigel, $\eta$ Carinae is believed to
be on the verge of exploding as a core-collapse supernova.

\item {\bf $\gamma^2$ Velorum} $\gamma^2$ Velorum is a binary consisting of a hot O-type star
and a Wolf-Rayet star. The proximity to the O-type star creates a situation where
two stellar winds interact, creating a wealth of
interesting phenomena such as wind collision zones, wind-blown
cavities and eclipses of spectral lines \cite{2007A&A...464..107M}.
The bright emission lines of WR stars make them suitable for
comparative studies in different passbands, as discussed above.

\end{itemize}

\paragraph{\bf Optical emission from AGN}
Nuclear optical continuum emission from AGN is visible whenever there is 
a direct view of the accretion disc (although jets can also
contribute to this component). NGC\,1068 is among the brightest and
the most nearby active galaxy (18.5\,Mpc), and hence the prototypical AGN. The core of
the galaxy is very luminous not only in the optical, but also in
ultraviolet and X-rays, and a supermassive black hole is required to
account for the nuclear activity. The VLTI has succeeded
in resolving structures in the AGN torus at mid-infrared 
wavelengths \cite{2004Natur.429...47J} on scales of 30 milli-arcsecond. On the other hand, 
the blue optical continuum emission
is dominated by thermal emission from the inner accretion disk (the
source of the ``big blue bump'' in many quasars) and much more
compact. The optical emission regions have an expected size of 0.3
milli-arcsecond at the distance of NGC\,1068 and resolving it would be
a fantastic achievement. The nucleus has a visual magnitude of around
$10^{\rm m}$.

\paragraph{\bf {Gamma-ray bursts and supernovae}}
The energy, spectrum and delay between the spectrally-remote events
and their inexplicable energy have long been puzzling.  Initial models
assumed clashes between expanding shells to explain these bright
events \cite{1998ApJ...497L..17S}, and later concentrated on polar shells.
Another model tried to explain the effect by assuming relativistic
plasma spheres in which directionality sets all these observed
parameters \cite{2003A&A...401..243D}.  Resolution is required to decide 
between the many models for these hot spots.
The currently available high-spatial resolution instruments
are all insufficient as amplitude interferometers cannot operate on
long enough baselines in the crucial blue and ultraviolet regime
\cite{2009ApJ...691..723B}. Early warning systems will allow to obtain some
spatial and spectral information at the peak of the optical flux. 
Since most of these events are extremely
far, we can expect only those within z$=1$ to be resolvable. There have been
at least two such bright ($\rm m_{v}<9^{\rm m}$) objects, GRB990123 and GRB080319B,
and there are others (e.g. GRB050509B) with host galaxies, all
of which are of interest.

\section{Cherenkov telescope arrays as an optical intensity interferometer}
\label{SIIandIACT} 
\begin{figure}
  \includegraphics[width=7cm]{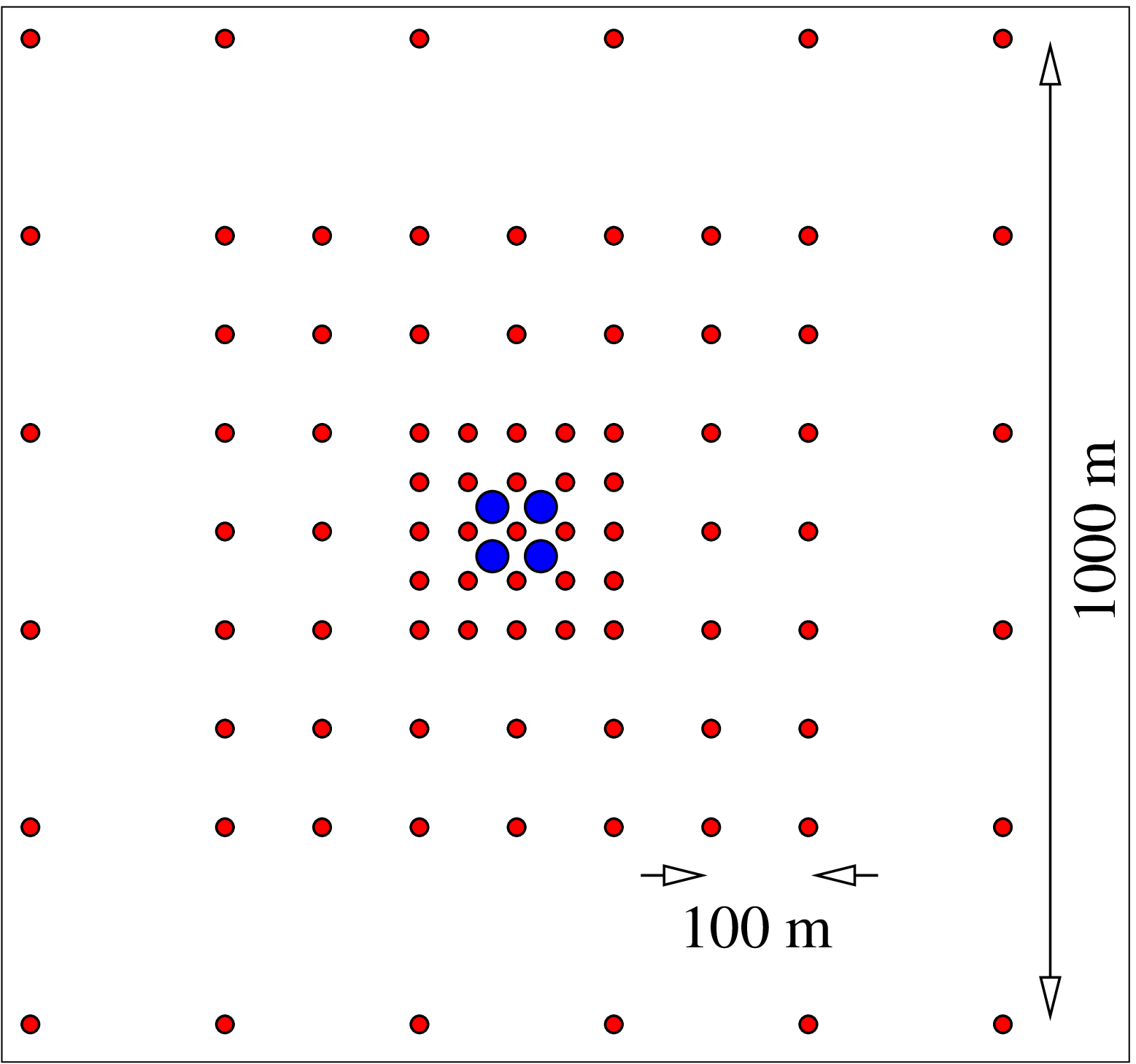}
  \includegraphics[width=9cm]{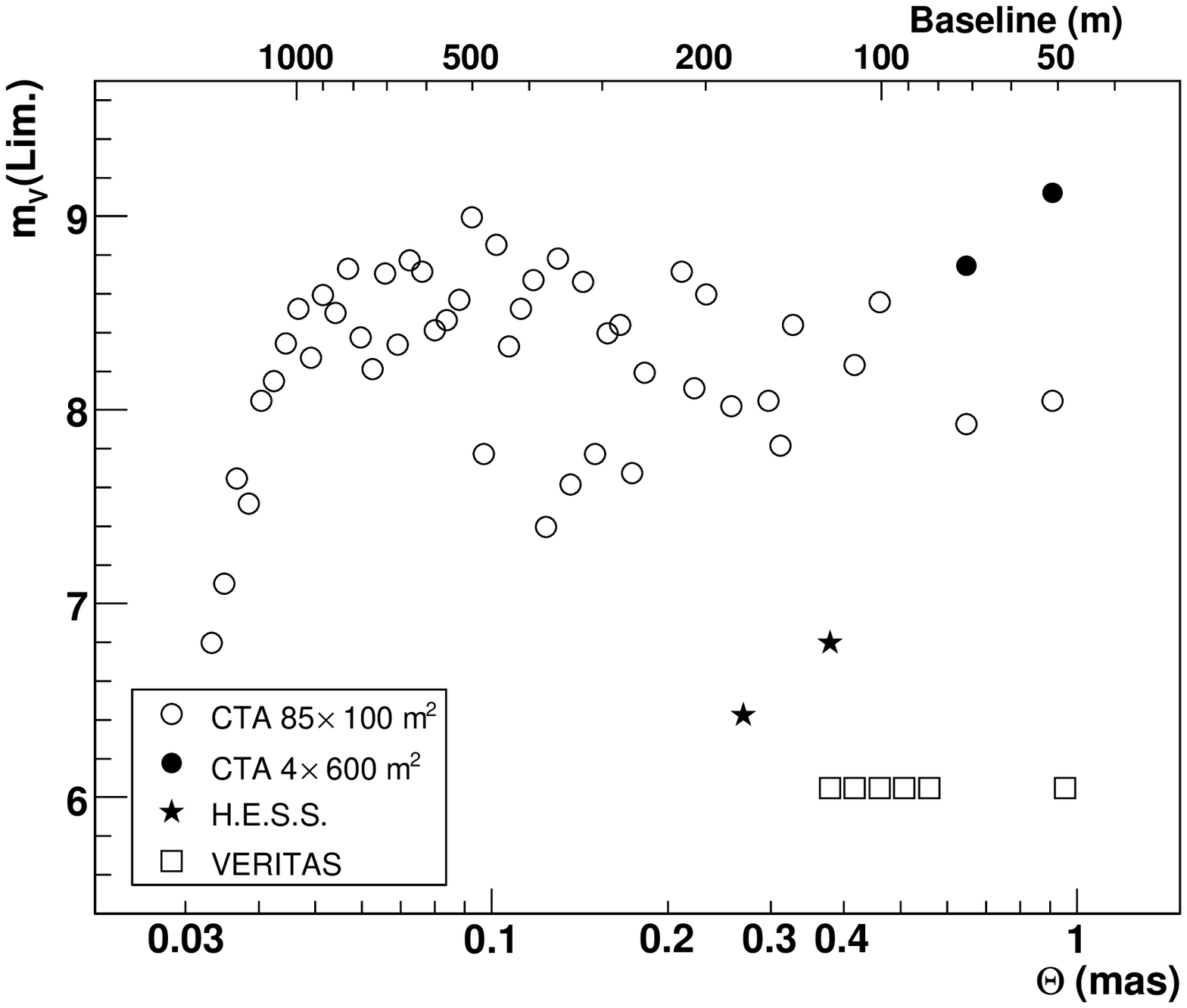}
  \caption{ {\it Left:} Proposed lay-out for the future CTA. Small
red dots are the 85 $\rm 100\,m^2$ dishes, large blue dots are the four $\rm
600\,m^2$ dishes (adapted from \cite{bernloehr}). {\it Right:} Limiting optical
magnitude estimate as function of baseline for a 5$\sigma$
detection, in a 5 hr integration on a centro-symmetric object with
50\% visibility. Sensitivity for $SII$ of current ground-based 
$\gamma$-ray facilities (HESS, and VERITAS) are also
included. Final limiting magnitudes depend on signal bandwidth and CTA design
details, see \cite{2006ApJ...649..399L}.}
  \label{figsiiiact}
\end{figure}

The major observational advantages of an intensity
interferometer are its low requirements on path length equalization and its
relative insensitivity to atmospheric seeing. However traditional 
implementation has required huge quantities of light. Cherenkov telescopes are
capable of gathering huge quantities of light and recent studies
\cite{2006ApJ...649..399L} have rediscovered the potential
of the next-generation Imaging Air Cherenkov Telescope (IACT) arrays as a multi-element
intensity interferometer. Two major IACT
array facilities for $\gamma$-ray astronomy are currently under design-study:
the Advanced Gamma Imaging System (AGIS \cite{agis2008}) in the US and 
the Cherenkov Telescope Array (CTA \cite{cta2008}) in Europe.
Current designs will offer several thousand baselines (tens to a hundred telescopes)
from a few tens of meters to more than a kilometer (Figure~\ref{figsiiiact}) and would achieve a
limiting magnitude for $SII$ of $\rm m_{v}\approx 9^{\rm m}$, achievable within a few
hours exposure \cite{2006ApJ...649..399L,
2008AIPC..984..205L,2008SPIE.7013E..72L}. CTA is on the
roadmap\footnote{\tt{ftp://ftp.cordis.europa.eu/pub/esfri/docs/esfri\_roadmap\_update\_2008.pdf}}
of the European Strategy Forum on Research Infrastructures (ESFRI), is stated
as one of the "Magnificent Seven" of the European strategy for astroparticle
physics published by
ASPERA\footnote{\tt{http://www.aspera-eu.org/images/stories/roadmap/aspera\_roadmap.pdf}}
and is highly ranked in the strategic plan for European astronomy of
ASTRONET\footnote{\tt{http://www.astronet-eu.org/}}. The AGIS collaboration has provided a white paper for the
Division of Astrophysics of the American Physical Society on the status and
future of ground-based TeV $\gamma$-ray astronomy \cite{2008arXiv0810.0444B}.
The next generation of IACT arrays are foreseen to come on-line within the next 10 years.
Implementation of $SII$ with IACT arrays will result in imaging capabilities
with an expected angular resolution of 0.05\,mas at 400\,nm. 

IACT $\gamma$-ray observations are done with a low duty cycle due to a
requirement for low night-sky background light levels excluding
moonlight.Meanwhile $SII$ observations in narrow optical bands are much less
affected by the background The $SII$ exploitation of large IACT arrays will
close to double their operational duty cycle for a modest additional cost. The
elegant combination of $SII-IACT$ will increase and diversify science output by
giving access to unprecedented imaging capabilities.

These new developments across the community have resulted in the
formation of a working group on the topic within the IAU commission 54
on optical and infrared
interferometry\footnote{\tt{http://physics.technion.ac.il/\,$\tilde{
}$\,intint/index.html}}.  The group met officially for the first time during a
workshop in January 2009\footnote{\tt{http://www.physics.utah.edu/\,$\tilde{
}$\,lebohec/SIIWGWS/}}. A report for the  CTA
collaboration is in preparation. Provisions will be included in the high energy
$\gamma$-ray camera design to make the future installation of $SII$
specific hardware possible.  The foreseen IACT arrays offer a sufficiently dense coverage of the
Fourier plane to perform actual image reconstruction. Implementation of $SII$ on IACT arrays is
identified as a natural first step towards revival of
intensity interferometry.


\section{Concluding remarks}
The natural synergy of IACT arrays and $SII$ will bring together
disparate research communities around a single large scale facility
with imaging capabilities at an unprecedented angular resolution. This
will usher in true imaging of diverse stellar phenomena such as rapidly
rotating stars and mass accretion processes. It will illuminate stellar
evolution processes through calibration of pre-main sequence tracks
and highly evolved systems close to going supernova. It will significantly
improve the distance scale by calibrating the Cepheid period-luminosity
relation.  Extra-galactic targets are fainter and smaller, but not out
of reach of future IACTs. The time for
revival of intensity interferometry in astronomy has arrived. 

The temporal variant of $SII$ has been chosen for a
planned quantum optics instrument for the European Extremely Large
Telescope, and is in an advanced stage of development. {\it Quanteye}
is designed to perform on sub-nanoseconds time scales allowing photon
correlation spectroscopy\cite{2006IAUS..232..506B,
2006SPIE.6269E..62N, 2007MSAIS..11..190B, 2006IAUS..232..502D}. It
will provide new insights into high-speed astrophysical phenomena and
the fine structure of photon emission. As astronomy is ultimately
driven by technical innovation rather than theoretical predictions,
these and other related developments of intensity interferometry 
will undoubtedly open up a new vantage point on the universe.




\section*{References}

\begin{scriptsize}

\providecommand{\newblock}{}

\end{scriptsize}

\end{document}